\documentclass[doublecol]{epl2} 
\usepackage{amsmath,amssymb,latexsym,MnSymbol}
\usepackage{hyperref}
\usepackage{color}
\usepackage{soul}
\usepackage[normalem]{ulem} 

\title{Long-range spatial correlations and fluctuation statistics of lightning activity rates in Brazil}

\author{H. V. Ribeiro\inst{1,2,3}\!\!\!\thanks{E-mail: \email{hvr@dfi.uem.br}} \and F. J. Antonio\inst{1,3} 
          \and L. G. A. Alves\inst{1,3} \and E. K. Lenzi\inst{1,3} \and R. S. Mendes\inst{1,3} }
\shortauthor{H. V. Ribeiro \etal}

\institute{                    
\inst{1} Departamento de F\'isica, Universidade Estadual de Maring\'a, Maring\'a, PR 87020-900, Brazil\\
\inst{2} Departamento de F\'isica, Universidade Tecnol\'ogica Federal do Paran\'a, Apucarana, PR 86812-460, Brazil\\
\inst{3} National Institute of Science and Technology for Complex Systems, CNPq, Rio de Janeiro, RJ 22290-180, Brazil\\
}


\abstract{
We report on a statistical analysis of the lightning activity rates in all Brazilian cities. We find out that the average of lightning activity rates exhibit a dependence on the latitude of the cities, displaying one peak around the Tropic of Capricorn and another one just before the Equator. We verify that the standard deviation of these rates is almost a constant function of the latitude and that the distribution of the fluctuations surrounding the average tendency is quite well described by a Gumbel distribution, which thus connects these rates to extreme processes. We also investigate the behavior of the lightning activity rates versus the longitude of the cities. For this case, the average rates exhibit an approximate plateau for a wide range of longitude values, the standard deviation is an approximate constant function of longitude, and the fluctuations are described by a Laplace distribution. We further characterize the spatial correlation of the lightning activity rates between pairs of cities, where our results show that the spatial correlation function decays very slowly with the distance between the cities and that for intermediate distances the correlation exhibits an approximate logarithmic decay. Finally, we propose to model this last behavior within the framework of the Edwards-Wilkinson equation.
}

\begin{document}
\maketitle

\section{Introduction}

The search for better understanding of earth-related systems is a stimulating challenge often present in the physicists agenda. Examples of such investigations include study of earthquakes~\cite{Mendes,Kawamura,Davidsen,Lippiello}, geomagnetic activities~\cite{Zheng,Reeves,Turner}, climate~\cite{Rybski,Boettle} and weather-related systems~\cite{Rybski2,Prahl}. Usually, these investigations examine data aiming to uncover patterns or laws that rule the spatiotemporal dynamics of the systems. These empirical-based approaches are also very useful to provide basis for comparing and testing the increasing number of models studied in these fields. 

In the previous context, a well-known phenomenon is the lightning strike, which displays an intrinsic complex behavior on its own. Physically speaking, lightning is a massive electric discharge between electrically charged regions in one or more clouds, or between clouds and an earth-bound object (cloud-to-ground lightning). Usually this complex process involves more than one current flow and lasts about half a second, while each one of these flows can take only milliseconds to happen~\cite{telesca2008}. Despite lightnings having short duration, they are well-known by causing severe injuries or even death when striking humans or animals, and for the damage they usually provoke to buildings and power networks~\cite{elsom2001, cooray2007}. In this sense, monitoring and probing spatiotemporal patterns of lightning occurrences may not only prevent electrical discharge hazard but also provide useful markers for thunderstorms, which are among the the major causes of weather-related damages and economic losses in tropical and middle latitude regions~\cite{enno2011}.

Around 78\% of all lightning events occur in tropical and middle latitude areas, that is, the majority of these events are concentrated between $30^\circ$ N and $30^\circ$ S latitudes~\cite{christian2003,williams2004}. Thus, understanding the lightning activity in these regions of the world is especially important. We devote this article to investigate the spatial dynamics of lightning activity rates in Brazil, the largest tropical country in the world, which is hit (on average) by more than 50 million lightning strikes each year. Furthermore, due its continental dimensions and the existence of reliable data, Brazil can be considered as an ideal place for probing spatial patterns of lightning activity. Here we shall study the lightning activity rates in connection with the geographic localization of all Brazilian cities aiming to uncover statistical patterns and to model these findings within a reductionist approach. In following, we present the data we have analyzed, the procedures we have employed to analyze the data as well as the empirical findings, and finally, we end with a summary and some concluding remarks.

\section{Data presentation}
\begin{figure}[!ht]
\centering
\includegraphics[scale=0.22]{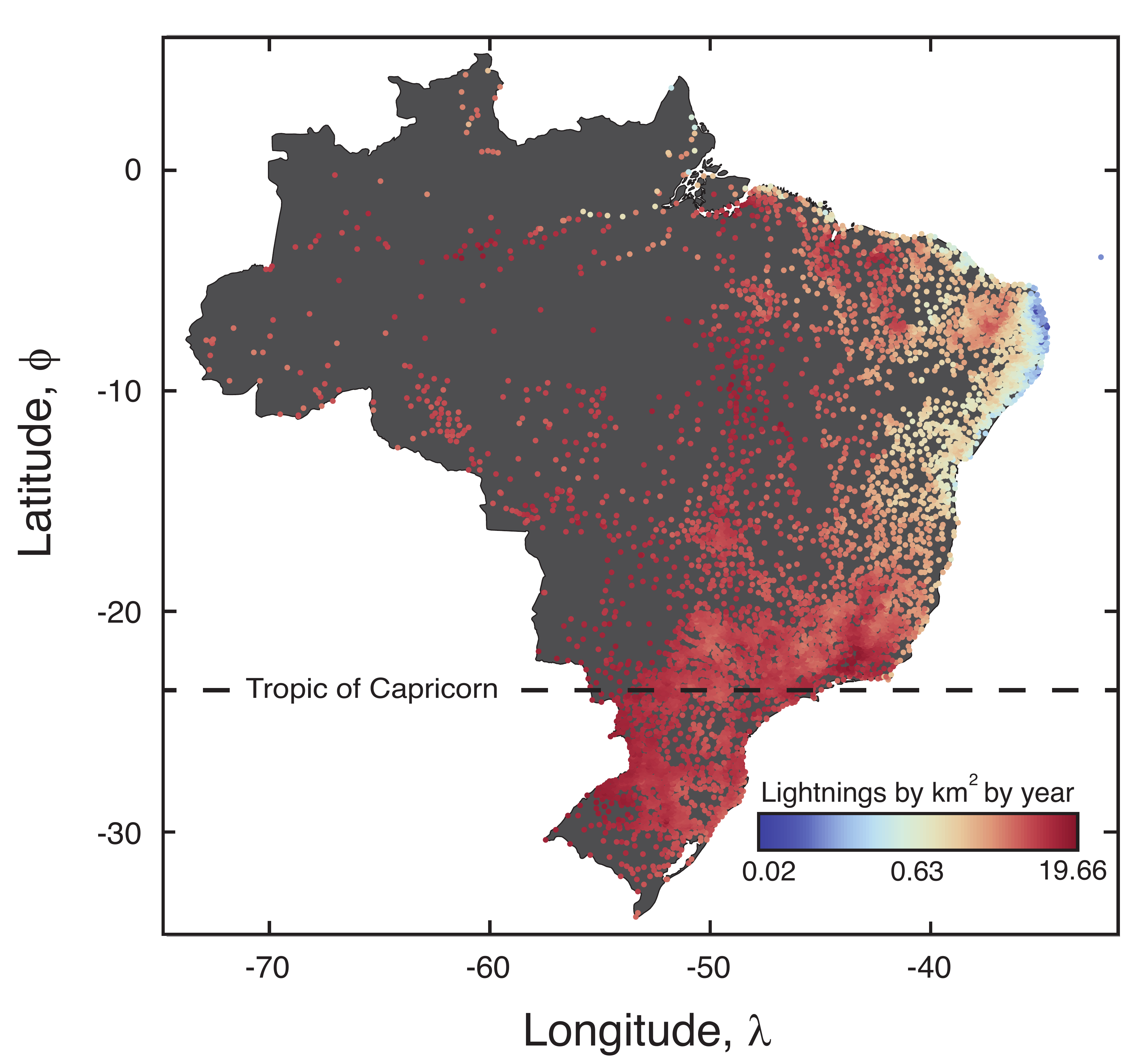}
\caption{{Spatial distribution of lightning activity rates of the Brazilian cities.} Each point represents the geographic location (latitude $\phi$ and longitude $\lambda$) of one of the 5,562 Brazilian cities and the color code indicates the lightning activity rate in the city.
}\label{fig1}
\end{figure}

We have accessed data corresponding to the lightning activity rates {(only cloud-to-ground lightning discharges)} of all Brazilian cities made freely available by the National Institute for Space Research (INPE)~\cite{ELAT,Pinto,Pinto2,Pinto3}. These rates were estimated by INPE through data obtained by the sensor Lightning Imaging Sensor~\cite{LIS} that is aboard the Tropical Rainfall Measuring Mission~\cite{TRMM}. The rates were calculated using data from the years 1998 to 2013 and are considered a reliable measure for all Brazilian cities. The data consist of the geographic location (latitude $\phi$ and longitude $\lambda$) and the average number of lightnings per km$^2$ per year ($L$). Figure~\ref{fig1} shows the spatial distribution of the lightning activity rates throughout the Brazilian territory. 

\section{Data Analysis}
\begin{figure*}[!ht]
\centering
\includegraphics[scale=0.46]{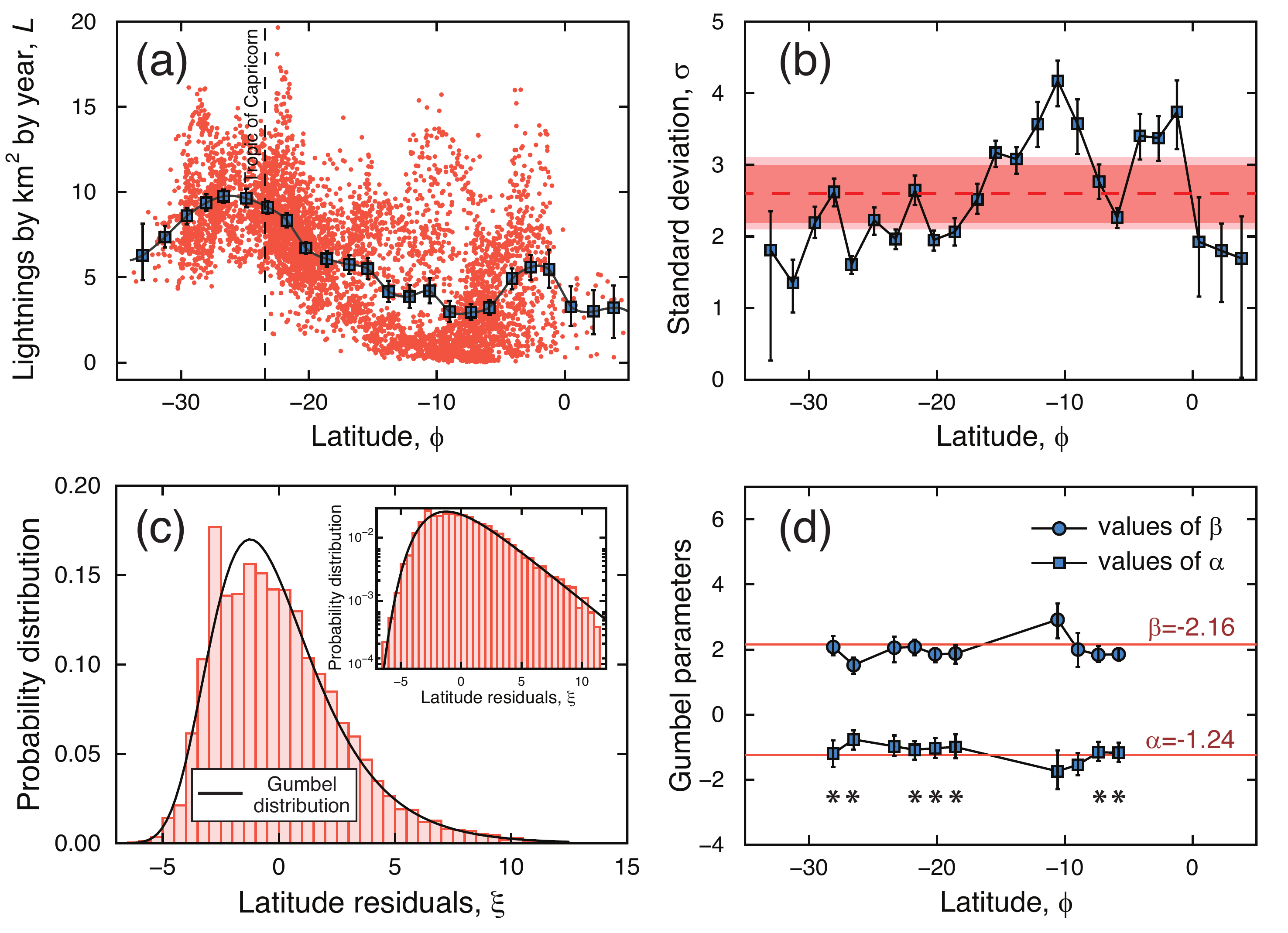}
\caption{{Dependence of the lightning activity rates on the latitude of the cities.} (a) Each red dot represents the lightning activity rate versus the latitude of a city. The blue squares are window average values (25 equally spaced windows) and the error bars are 95\% bootstrap confidence intervals for these means. The continuous line is a spline interpolation of third order to the average data. We note that the average value displays a maximum value close to the Tropic of Capricorn (horizontal dashed line) and also a local maximum just before the Equator (zero latitude). (b) Standard deviation of the lightning activity rates versus the latitude evaluated over same windows utilized for the average values. The blue squares are the standard deviations and the error bars are 95\% bootstrap confidence intervals. The horizontal dashed line is the average value of standard deviations and the dark (light) shaded area represents the 95\% (99\%) bootstrap confidence intervals. We note that except for a few points, the standard deviations do not display statistically significant deviations from a constant behavior. (c) Probability distribution of the residuals of the lightning activity rates around the average tendency described by the spline interpolation. The bins were chosen through the Wand's procedure and the inset shows the empirical distribution in log-log scale. The continuous line is the Gumbel distribution (eq.~\ref{eq:gumbel}) with location parameter $\alpha=-1.24$ and scale parameter $\beta=2.16$, which were obtained via maximum likelihood fit. The $p$-value of Cram\'er-von Mises test is 0.02. { (d) Values of $\alpha$ and $\beta$ of the Gumbel distribution when grouping the residuals by using the same 25 windows employed for the calculation the averages and standard deviations. These values were obtained via maximum likelihood fit considering only the windows with more than 250 points within and the asterisks indicate the windows where the Gumbel hypothesis cannot be rejected at a confidence level of 99$\%$. The error bars are 95\% bootstrap confidence intervals and the horizontal lines represent the $\alpha$ and $\beta$ obtained for the whole dataset.}
}\label{fig2}
\end{figure*}

We start by investigating the dependence of the lightning activity rates $L$ on the latitudes $\phi$ of the cities. Figure~\ref{fig2}(a) shows a scatter plot of $L$ versus $\phi$ where, despite the considerable noise, we can identify some regularities. We observe the existence of systemically large values for the lightning activity rates around the Tropic of Capricorn and also around the Equator. In order to check whether these visual patterns are statistically meaningful, we have binned the data into 25 windows and, for each window, we calculate the average value (square) of the lightning activity rates as well as the 95\% bootstrap confidence interval (error bar). As shown in fig.~\ref{fig2}(a), the average value of $L$ is actually statistically significantly larger around the Tropic of Capricorn and the Equator. We note yet the existence of maximum values just before these geographic references.

We have also studied the dependence of the standard deviation of the lightning activity rates on the latitude. Figure~\ref{fig2}(b) reveals that the standard deviation $\sigma$ is almost a constant function of the latitude. In particular, we note that practically all values of $\sigma$ are within the 95\% confidence interval (light shaded area). A systematic exception occurs around the latitude $-10$, where the standard deviations remain outside of the 99\% confidence interval for a constant behavior (dark shaded area). 

Aiming to further characterize the lightning activity rates, we have evaluated the residuals surrounding the average tendency of $L$ versus $\phi$. In order to do so, we have interpolated a spline of third order to the average data and then we calculate the residuals
\begin{equation}\label{eq:res}
\xi_i=L_i-F(\phi_i)\,,
\end{equation}
where $L_i$ is the lightning activity rate in the city $i$, $\phi_i$ is latitude of the city $i$ and $F(\phi_i)$ stands for the value of the interpolating function at the latitude $\phi_i$. Next, we investigate the probability distribution of $\xi$, as shown in fig.~\ref{fig2}(c). We have found that the Gumbel distribution~
\cite{Gumbel}
\begin{equation}\label{eq:gumbel}
P(\xi) = \frac{1}{\beta}\,e^{(-\xi+\alpha)/\beta-e^{(-\xi+\alpha)/\beta}}
\end{equation}
with the parameters $\alpha = -1.24$ (location parameter) and $\beta = 2.16$ (scale parameter) describes quite well our empirical distribution. The best fit parameters were obtained by the maximum likelihood method and the $p$-value of the Cram\'er-von Mises test is 0.02, indicating that we cannot reject the Gumbel hypothesis at a confidence level of 99\%. 

{ In order to overcome possible bias related to the inhomogeneous spatial distribution of the Brazilian cities, we have also calculated the distribution of $\xi$ after grouping cities with close latitudes values, considering the same 25 windows employed for the average and standard deviation analyses (figs.~\ref{fig2}(a) and (b)). The results show that the distributions of the residuals by latitude obey the same Gumbel form with parameters $\alpha$ and $\beta$ very close to those obtained for the whole dataset, as shown in fig.~\ref{fig2}(d).} It is worth noting that the Gumbel distribution is related to the distribution of the maximum of a set of $n\to\infty$ random numbers drawn from a distribution that asymptotically decays faster than any power law. This result thus suggests that the deviations surrounding the average tendency can be also understood as an extreme process. { While it is not trivial to connect these results with the mechanisms underlying the physics of lightning process~\cite{Rakov}, we can consider a cloud-to-ground lightning as an extreme event of the several processes that occur in a thundercloud (in addition to large values of the electric fields necessary for the lightning initiation). Actually, cloud-to-ground lightnings are about 25$\%$ of the global lightning activity, all remaining activity does not involve the ground and it is mainly composed by intra-cloud discharges and by (on a much smaller fraction) cloud-to-cloud and cloud-to-air discharges. Therefore, when analyzing only cloud-to-ground lightnings, we are selecting a group of extreme events and thus approaching somehow the mathematical conditions of the extreme value theory where Gumbel distributions naturally emerge.} 

\begin{figure*}[!ht]
\centering
\includegraphics[scale=0.46]{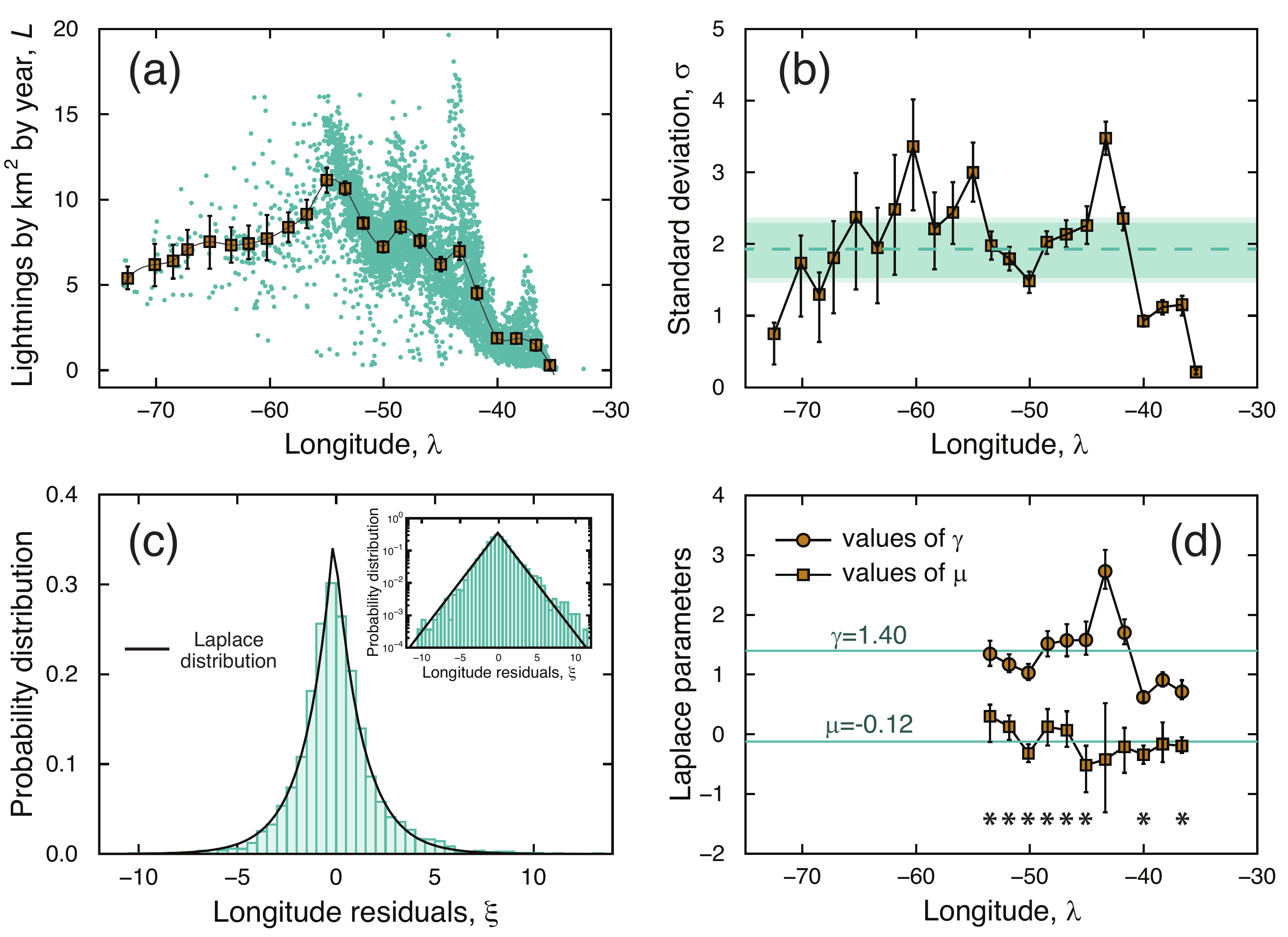}
\caption{{Dependence of the lightning activity rates on the longitude of the cities.} (a) Each green dot represents the lightning activity rate versus the longitude of a city. The orange squares are window average values (25 equally spaced windows) and the error bars are 95\% bootstrap confidence intervals for these means . The continuous line is a spline interpolation of third order to the average data. We note that the profile of the average values is almost a plateau in the interval $-70\lesssim\lambda\lesssim-43$ and also that the lightning activity rates are substantially smaller for longitudes larger than $\approx-40$. (b) Standard deviation of the lightning activity rates versus the longitude evaluated over same windows utilized for the average values. The orange square are the standard deviations and the error bars are 95\% bootstrap confidence intervals. The horizontal dashed line is the average value of standard deviations and the dark (light) shaded area represents the 95\% (99\%) bootstrap confidence intervals. Similarly to the latitude case, the standard deviations do not display statistically significant deviations from a constant behavior. (c) Probability distribution of residuals of the lightning activity rates around the average tendency described by the spline interpolation. The bins were chosen through the Wand's procedure and the inset shows the empirical distribution in log-log scale. In contrast with the latitude case, the continuous line is a Laplace distribution (eq.~\ref{eq:laplace}) with mean $\mu=-0.12$ and scale parameter $\gamma=1.40$, which were obtained via maximum likelihood fit. The $p$-value of Cram\'er-von Mises test is 0.02. { (d) Values of $\mu$ and $\gamma$ of the Laplace distribution when grouping the residuals by using the same 25 windows employed for the calculation the averages and standard deviations. These values were obtained via maximum likelihood fit considering only the windows with more than 250 points within and the asterisks indicate the windows where the Laplace hypothesis cannot be rejected at a confidence level of 99$\%$. The error bars are 95\% bootstrap confidence intervals and the horizontal lines represent the $\mu$ and $\gamma$ obtained for the whole dataset.}
}\label{fig3}
\end{figure*}

We now focus on the dependence of the lightning activity rates $L$ over the longitude $\lambda$ of the cities. We have proceeded as previously, that is, we start with a scatter plot of $L$ versus $\lambda$, we bin the data into 25 windows and for each one we evaluate the average of $L$, and we also interpolate a spline of third order to the average data, as it is shown in fig.~\ref{fig3}(a). For this case, we do not observe a clear dependence of the average lightning activity rates over the longitudes within the interval $-70\lesssim\lambda\lesssim-43$. In this interval, apart from a peak around the latitude $-55$, $L$ displays an approximate constant behavior. However, we do note a systematic decrease in the average values of $L$ for for longitudes higher than $\approx-43$. This fact is also visible in fig.~\ref{fig1}, where the blue region points out the lower lighting rates in the northeast of Brazil. Analogously to the previous case, we have also studied the standard deviations $\sigma$ within windows in function of the longitude (fig.~\ref{fig3}(b)). Similarly, we have found an almost constant behavior and that most of the values of $\sigma$ are inside the 99\% confidence intervals of the average of the standard deviation over the longitude (though systematic deviations are observed for longitudes larger than $-43$).

The most striking difference between the relationships of $L$ versus latitude and $L$ versus longitude occurs for the distribution of the residuals $\xi$ around the average tendency. While the Gumbel distribution is a quite good description for the dependence of $L$ on the latitude, it cannot describe the dependence of $L$ on the longitude. In fact, the tent-shaped distribution shown in fig.~\ref{fig3}(c) is in well agreement with the Laplace distribution
\begin{equation}\label{eq:laplace}
P(\xi) = \frac{1}{2 \gamma}\,e^{-|\xi-\mu|/\gamma}\,,
\end{equation}
where the parameters $\mu=-0.12$ (mean) and $\gamma=1.40$ (scale factor) were obtained via maximum likelihood fit and the $p$-value of Cram\'er-von Mises test is 0.02, pointing out that the Laplace hypothesis cannot be rejected at a confidence level of 99\%. { As in the latitude case, we also calculated the distribution of $\xi$ after grouping cities with close longitudes values. Intriguingly, the Laplace form also describes the residuals distributions by longitude with parameters $\mu$ and $\gamma$ very close to those obtained for the whole dataset, as shown in fig.~\ref{fig3}(d). In fact, since the spatial lightning distribution is closely linked to climate phenomena on Earth (such as the general circulation of the atmosphere between the equator and the middle latitude~\cite{Price}) and therefore it seems unrelated to longitudes, we could expect normal distributions emerging after accounting the bias related to the inhomogeneous spatial distribution of the Brazilian cities. However, the lightning rates are also influenced by the continental and oceanic forms of the Earth, which introduce a non-trivial dependence of the lightning rates on longitudes~\cite{Price2}.} 

\begin{figure}[!ht]
\centering
\includegraphics[scale=0.46]{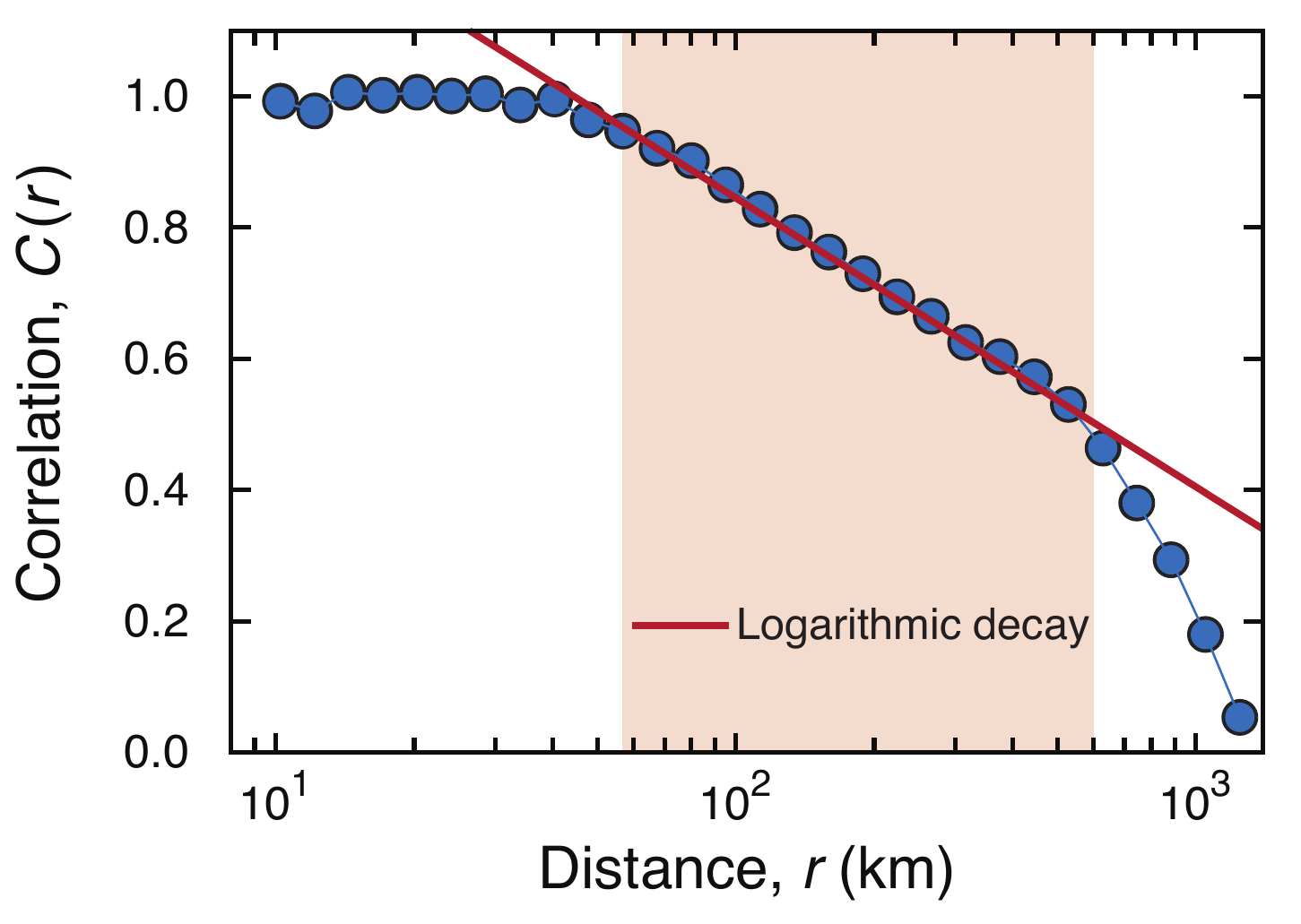}
\caption{{Slow decay of the spatial correlations in lightning activity rates.} Dependence of the spatial correlation function $C(r)$ on the distance $r$ (circles). We note that for small distances ($r\lesssim40\,\text{km}$) the correlation remains almost constant. We next observe an approximate logarithmic decay, where $C(r)\approx-\mathcal C \ln(r/\mathcal R)$ describes the correlation function for $60\,\text{km}\lesssim r\lesssim600\,\text{km}$ (shaded area) with ${\mathcal C}=0.19$ and ${\mathcal R}=8311.32$. For $r\gtrsim600\,\text{km}$, we note a faster-than-logarithm decay.
}\label{fig4}
\end{figure}

Another intriguing question is whether there is long-range spatial memory in the lightning activity rates $L$. To investigate this hypothesis, we evaluate the spatial correlation function of the lightning activity rates between pairs of cities that are $r$ kilometers distant. Specifically, we have computed 
\begin{equation}\label{eq:corr}
C(r) = \frac{\langle [L_i - m(r)] [L_j - m(r)] \rangle\vert_{r_{i,j}=r}}{{s(r)}^2}\,,
\end{equation}
where $m(r)$ is the mean value and $s(r)$ is the standard deviation of the lightning activity rates of cities separated by $r$ kilometers, $L_i$ is the lightning activity rate in the city $i$, and $\langle \dots \rangle\vert_{r_{i,j}=r}$ stands for the average value over cities where the distance $r_{i,j}$ that separates them are equal to $r$. Due to the discrete nature of our spatial data, we have actually considered logarithmically spaced intervals of $r$ for evaluating eq.~\ref{eq:corr}. Figure~\ref{fig4} shows the spatial correlation $C(r)$ in a lin-log scale where it is remarkable a very slow decay of the correlation function. In particular, we have identified three regimes: $i)$ an initial plateau in the interval $10\,\text{km} \lesssim r\lesssim40\,\text{km}$ where $C(r)\approx1$; $ii)$ an approximate logarithmic decay in the interval $60\,\text{km} \lesssim r\lesssim600\,\text{km}$ (shaded area) where $C(r)\approx-\mathcal C \ln(r/\mathcal R)$; and $iii)$ a faster-than-logarithm decay for distances larger than $\approx600\,\text{km}$. We thus conclude that the lightning activity rates are long-range correlated in space, since the correlation function $C(r)$ decays slower than any power law function for distances smaller than $\approx600\,\text{km}$. { Another interesting point is concerning the relationships between these long-range correlations and the non-Gaussian residual distributions previously presented. In several contexts, long-range correlations emerge in coupled-like manner with non-Gaussian distributions (see Ref.~\cite{Mendes2} for a specific example) and here the same happens. Furthermore, if we shuffle the lightning rates among the cities, besides destroying the spatial correlations this process drastically changes the profile of the residual distributions. Specifically, both residual distributions for latitude and longitude become equal and they pass simply to reflect the spatial distribution of the Brazilian cities.}

The approximate logarithmic decay is very intriguing and it can be modeled on the theoretical canvas of the Edwards-Wilkinson equation~\cite{Edwards_Wilkinson}
\begin{equation}\label{eq_EW}
\frac{\partial }{\partial t}L({\bf r},t) = D\, \nabla^2 L({\bf r},t) + \eta({\bf r},t)\,,
\end{equation}
where $L({\bf r},t)$ represents the lighting rate in a point localized by the vector ${\bf r}$, $D$ is a constant, $\nabla^2$ is the Laplacian, and $\eta({\bf r},t)$ is an uncorrelated noise (in space and time) with zero mean and finite variance. { Because this equation is mainly used to describe stochastic kinetics of a growing interface, it may appear as an \textit{ad hoc} hypothesis for modeling our data. However, the Edwards-Wilkinson equation contains two important phenomenological ingredients related to the spatial distribution of the lightning activity: diffusion and randomness. The former (Laplacian term) can be related to the diffusive aspects of the thunderstorms and the second (noise term) reflects the natural complexity of the lightning phenomenon (as well as other sources of randomness). Nonetheless, modeling the spatial distribution of lighting rates only with this equation represents a quite crude approximation where several important processes are not considered. For instance, we can not expect this simple model to describe the residual distributions. On the other hand,} the autocorrelation function related to eq.~\ref{eq_EW} can be obtained in $d$ dimensions by taking Fourier transforms in space and time, evaluating the correlation in this double-transformed Fourier space, and next returning to the usual space via inverse Fourier transforms (see Refs.~\cite{Edwards_Wilkinson,Nattermann,Stanley} for more details). For equal-times and in the equilibrium ($t \gg {\mathcal R}^2/D$, see below), the correlation is written as $C(r)\sim r^{2-d}$ when $d\neq2$, and for the two-dimensional case (that is, for our case) it is
\begin{equation}
C(r) \approx - {\mathcal C} \ln \left(\frac{r}{{\mathcal R}}\right)\quad \text{for}\quad r_{\text{min}}<r<{\mathcal R}\,,
\end{equation}
where ${\mathcal C}>0$ is constant, $r_{\text{min}}$ is a short-distance cut-off and ${\mathcal R}$ is the length of the system. We thus observe that the logarithm decay is intimately connected with the two-dimensional nature of the problem. Furthermore, it is equally intriguing that similar logarithm decays were reported in very different context related to turnout rates in elections~\cite{Borghesi,Borghesi2} { and spatial spreading of obesity and other diseases~\cite{Gallos}}, which can be somehow related to remnant effects of a universal mechanism underlying these two-dimensional processes. 

\section{Summary}
We characterized the lightning activity rates over all the Brazilian cities by taking their geographic location into account. We firstly studied the dependence of the lightning activity rates on the latitude of the cities, where we observed two peaks in the average value of these rates: one just before the Tropic of Capricorn and another one just before the Equator. We also investigated the fluctuations surrounding the average tendency, where we found that the standard deviation is almost a constant function of the latitude and that the distribution of these fluctuations (residuals) is in quite good agreement with a Gumbel distribution. The dependence of the lightning activity rates on the longitude of the cities was also characterized, where we reported that the average value exhibits an approximate plateau for a wide range of longitude values. For the longitude case, the standard deviation of the residuals also presents an almost constant behavior; however, the distribution of the residuals was well described by a Laplace distribution. We further investigated the spatial correlations of the lightning activity rates between a pair of cities. Our results show that the correlation function decays very slowly with the distance between the cities and that for intermediate distances ($60\,\text{km}\lesssim r\lesssim600\,\text{km}$) the correlation exhibits an approximate logarithm decay. Finally, we proposed to model the logarithmic decay via the Edwards-Wilkinson equation, where we observed that this behavior is closely related to two-dimensional nature of the system and also that the same behavior was reported in a very different context, which somehow points to a universal mechanism.

\acknowledgments
We thank Capes, CNPq and Funda\c{c}\~ao Arauc\'aria for financial support. H.V.R. is especially grateful to Capes/Funda\c{c}\~ao Arauc\'aria for financial support under grant \mbox{number 113/2013}.

\end{document}